\documentclass{vgtc}                          

\graphicspath{{figures/}{pictures/}{images/}{./}} 

\usepackage{times}                     

\usepackage{tabu}                      
\usepackage{booktabs}                  
\usepackage{lipsum}                    
\usepackage{mwe}                       

\usepackage{balance}
\usepackage{soul}

\usepackage{mathptmx}                  

\onlineid{0}

\vgtccategory{Research}

\vgtcinsertpkg

\newcommand{\changed}[1]{\textcolor{black}{#1}}

\title{Extended Reality as a Mediation Layer for Situated Human Control in Human-Robot Teaming}

\author{Jens Grubert\thanks{e-mail: jens.grubert@hs-coburg.de}\\ %
        \scriptsize Coburg University of Applied Sciences and Arts %
\and John Dudley\thanks{e-mail: jjd50@cam.ac.uk}\\ %
     \scriptsize University of Cambridge %
\and Eyal Ofek\thanks{e-mail: e.ofek@bham.ac.uk}\\ %
     \scriptsize University of Birmingham %
\and Per Ola Kristensson\thanks{e-mail: pok21@cam.ac.uk}\\ %
     \parbox{1.4in}{\scriptsize \centering University of Cambridge}}

\abstract{

Extended Reality (XR) is increasingly used in human-robot interaction
to communicate robot intent, planned motion, reachability, and state.
We argue that XR should also be understood as a
mediation layer for situated human control in human-robot teaming.
Situated human control denotes the human collaborator's ability to
understand, shape, authorize, and interrupt robot action within the
concrete physical, social, and temporal context in which that action
unfolds. We ground this perspective in scenarios from robot-assisted
bedside nursing, multi-arm supervisory control, and collaborative
assembly under divided attention. Across these scenarios, robot autonomy
must remain inspectable and adjustable as people move, goals change,
sensing is incomplete, control roles shift, and plans become invalid.
We identify four mediation functions connecting human intent and robot
autonomy, robot plans and human judgment, levels of shared control, and
team roles, handover, and recovery. Building on these functions, we derive six design
dimensions: joint action possibilities, socio-physical constraints,
uncertainty and plan validity, multimodal control and correction, roles,
handover, and accountability, and anticipatory recovery. The paper
outlines a research agenda for XR systems that make robot autonomy more
actionable and accountable in dynamic shared environments.

}


\begin{document}


\firstsection{Introduction}

\maketitle

Human-robot interaction (HRI) is increasingly moving toward teams in which people collaborate with autonomous, semi-autonomous, or teleoperated robots in shared physical environments. \changed{Coordination in
such settings requires reciprocal anticipation: people need to
understand robot behavior, while robots must account for human actions,
attention, and task state.} 
This becomes especially important in dynamic environments where people move, goals change, sensing is incomplete, and previously valid robot plans may become inappropriate during execution.


Extended Reality (XR) is well suited for this challenge because it can present robot state, intent, constraints, and interaction possibilities directly in the workspace where robot actions unfold \cite{suzuki2022augmented,walker2023vam,pascher2023motionintent}. Prior work has shown that XR can support HRI by visualizing planned motion, reachability, robot state, task progress, and authorable robot behaviors \cite{gruenefeld2020mindthearm,hauck2025reachvox,lunding2025arthur}. These approaches demonstrate the value of making robot behavior spatially legible. However, emerging HRI scenarios require interfaces that go beyond awareness of robot intent. Human collaborators may need to assess whether a proposed action remains appropriate, constrain or correct a plan, approve execution, or interrupt action when the situation changes.



\changed{XR devices can also contribute information about the human collaborator.
Head pose, gaze, hand motion, and explicit gestures can provide spatial
and deictic cues that help robots interpret references, estimate
attention, or detect requests for intervention. Such signals should not
be treated as direct or reliable measurements of human intent; instead,
they provide additional, uncertain evidence that can be combined with
task context and explicit input \cite{si2023multimodal,lunding2025arthur}. This bidirectional role distinguishes
XR mediation from interfaces that only display robot-generated
information.}

We argue that XR should be conceptualized as a mediation layer for \emph{situated human control} in human-robot teaming. We use situated human control to denote the human collaborator's ability to understand, shape, authorize, and interrupt robot action within the concrete physical, social, and temporal context in which that action unfolds. This notion of control is broader than direct teleoperation: it includes supervisory control, constraint specification, plan approval, role handover, and recovery from emerging misalignments. XR can support such control by making robot autonomy visible in context and by turning human judgment into actionable corrections, constraints, approvals, or interruptions.

This framing is motivated by two recurring HRI challenges. First, robot plans are often technically feasible while remaining questionable in context. A trajectory may avoid collisions but still block a worker's access, pass too close to a vulnerable patient, violate social expectations, or depend on uncertain sensing. Second, human input is increasingly high-level and multimodal. Speech, gaze, and gesture allow efficient task specification, but they also introduce ambiguity about referents, intent, and control scope \cite{si2023multimodal,lunding2025arthur}. In multi-person settings, such as care teams or supervised training, these problems are further complicated by role changes and handover of control.


\changed{This paper makes three contributions. First, we define XR
mediation as a bidirectional, situated coupling between human or team
state, robot autonomy, and the shared environment. Second, we use
three contrasting HRI scenarios to identify recurring coordination
breakdowns involving spatial consequences, ambiguous input, changing
plan validity, divided control, and recovery. Third, we derive six scenario-grounded design dimensions and
associated research and evaluation considerations for XR interfaces
that support situated human control.}


\section{Motivating Scenarios}

\begin{figure*}  
    \centering
    \includegraphics[width=\linewidth]{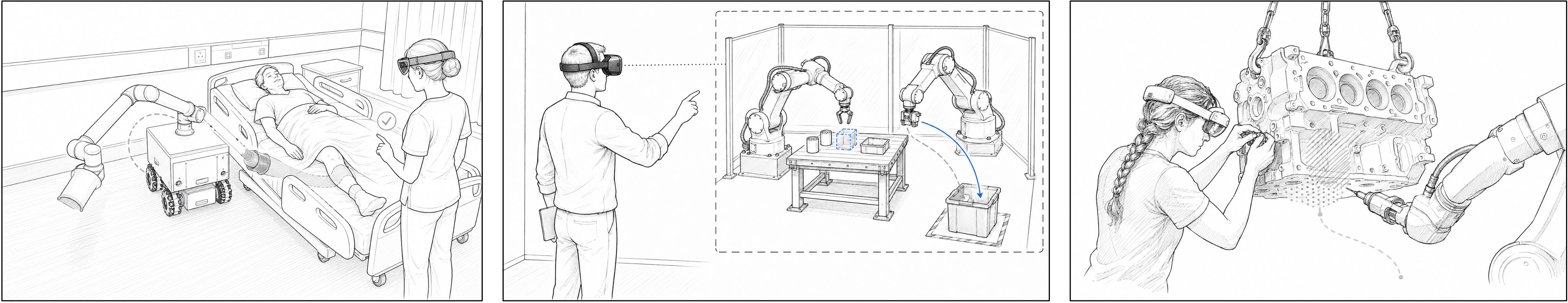}
    \caption{Illustrations of the three motivating scenarios. Left: robot-assisted bedside nursing. Center: multi-arm supervisory control. Right: collaborative assembly.}
    \label{fig:scenarios}
\end{figure*}

We ground our argument in three scenarios that differ in domain, robot configuration, and interaction style, but share a common design problem: robot autonomy must remain understandable and actionable as the situation changes, see Figure \ref{fig:scenarios}. In each scenario, XR can support situated human control by making planned robot action visible in context and by enabling humans to correct, constrain, approve, interrupt, or recover that action.

\subsection{Robot-Assisted Bedside Nursing}

In robot-assisted bedside nursing, a collaborative robot may support physically demanding care tasks such as lifting, stabilizing, or repositioning a patient's limb during wound care, hygiene, or preparation for a dressing change \cite{grubert2026mapple}. Such tasks can impose high musculoskeletal strain on nurses and may currently require a second caregiver. Reviews of nursing robotics highlight both the potential of assistive robots and the need for stronger nurse-centered design and evaluation in realistic care contexts \cite{soriano2022robots,ohneberg2023assistive,babalola2024collaborative}.

This scenario illustrates why robot intent must be interpreted relative to the care situation. A planned robot motion may be collision-free while still being unacceptable because it passes close to the patient's face, obstructs the nurse's access to the treatment site, approaches an injured region, or causes discomfort. XR can mediate this interaction by previewing the planned motion relative to the patient's body, showing expected contact or support regions, indicating uncertainty in patient tracking, and allowing the nurse to approve, constrain, delay, or interrupt execution. Situated human control is grounded here in caregiver authority: the nurse remains responsible for the care activity, while the robot contributes physical support.


The scenario also foregrounds multi-person role and control management. Bedside care may involve nurses, physicians, trainees, patients, and relatives. The robot must distinguish authorized commands from incidental conversation, teaching explanations, or patient speech. XR can support this by making the currently authorized operator, the scope of control, pending approvals, and available override mechanisms visible to the care team.

\subsection{Multi-Arm Supervisory Control}

A second scenario concerns a teleoperator coordinating multiple robot arms in a remote or hazardous workspace. Directly controlling each arm can be inefficient because it forces the operator to serialize attention and low-level control. A more scalable strategy is supervisory control, where the operator issues high-level instructions and intervenes when ambiguity, risk, or task complexity requires closer involvement \cite{stotko2019vr,sheridan1992telerobotics,si2023multimodal}.

XR is well suited to this setting because it can present the remote workspace spatially while supporting speech, gaze, gesture, and direct manipulation. For example, an operator might look at an object, point to a target location, and say ``move this to the left bin,'' while the system infers the object, target, and responsible robot arm. Such input is efficient, but it creates ambiguity: which object was selected, which arm should act, which path will be used, and whether another arm or human access space will be affected. XR can externalize these interpretations before execution.

This scenario highlights transitions along an intent-to-precision continuum. The operator may begin with high-level intent, refine the plan through deictic constraints, directly adjust a virtual path, or temporarily take fine-grained control. Situated human control depends on the interface making these transitions legible: the user should understand when they are supervising autonomous execution, when they are specifying constraints, and when they are directly controlling motion.

\subsection{Collaborative Assembly Under Divided Attention}

A third scenario concerns collaborative assembly in which a human and robot work on interdependent but partially parallel subtasks. For example, a robot may fetch a tool, hold a component, or prepare the next object while the human completes a manual step. Unlike bedside nursing, where the caregiver remains focused on patient care and robot action near the body, this scenario emphasizes divided attention: the human may be occupied with their own subtask and may only intermittently monitor the robot.

This creates a different challenge for situated human control. Small mismatches between robot action and task progress can cascade into larger failures. The robot may bring the wrong part, approach before the human is ready, block access to the next work area, or continue executing a plan that no longer matches the human's current sequence. XR can support anticipatory recovery by visualizing future robot actions, timing conflicts, and likely interference points before they become task failures.

In this scenario, XR should support low-effort intervention under cognitive load. Planned robot actions can be represented as manipulable spatial objects: a target can be reassigned, a timing conflict can be paused, a path can be shifted away from the human's workspace, or an action can be redirected to a safer alternative. This connects robot intent visualization with direct manipulation and shared control \cite{pascher2023motionintent,gruenefeld2020mindthearm,dragan2013policy}. The key question is whether XR allows users to notice and correct emerging misalignments while maintaining continuity in their own task.


\section{XR as a Mediation Layer}

\changed{The scenarios above illustrate a common problem: robot actions may be
technically feasible while remaining inappropriate in the concrete
situation in which they will unfold. Human collaborators therefore need
more than awareness of robot intent. They must be able to relate a
robot's interpretation and planned behavior to the current task, people,
workspace, uncertainty, and distribution of control, and to intervene
when these relations become misaligned.}

\changed{Intent visualization remains an important foundation for XR-based HRI.
Planned paths, goals, reachability, state, and task progress can help
users anticipate robot behavior and coordinate with it
\cite{pascher2023motionintent,suzuki2022augmented,walker2023vam,
gruenefeld2020mindthearm,hauck2025reachvox}. Situated human control
extends this role from awareness to actionability. An XR interface should
not only communicate what the robot plans to do, but also help users
assess situated consequences, understand the current control relation,
compare alternatives, and correct, approve, interrupt, hand over, or
recover robot action.}

\changed{We conceptualize the XR mediation layer as a bidirectional coupling
between human or team state, robot autonomy, and the shared environment.
Robot-generated information flows toward human collaborators through
spatial representations of interpretations, plans, uncertainty, control
modes, and anticipated consequences. Human input flows toward the robot
through speech, gaze, gesture, direct manipulation, constraints, role
assignment, approval, and interruption. The shared environment grounds
both directions by relating input and robot behavior to concrete people,
objects, regions, tasks, and events. Mediation therefore comprises
contextualization and disambiguation as well as visualization,
intervention, and recovery.}

\changed{XR is not uniquely required for every mediation function. Conventional
displays can communicate robot state, uncertainty, or control modes. XR
becomes particularly advantageous when mediation depends on spatial
registration, embodied and deictic input, mobility, divided attention, or
a shared view of physically situated action. It can anchor plans and
constraints to robots, people, objects, and regions in the workspace;
capture gaze, head, and hand activity as spatial input; and support local
or remote users without requiring repeated attention shifts to a separate
display. The relevant design question is therefore not whether XR can
display a particular item of information, but whether spatial grounding
and embodied interaction improve the human's ability to understand or
influence robot action
\cite{suzuki2022augmented,walker2023vam,si2023multimodal}.}

\changed{The following subsections distinguish four mediation functions:
mediating between human intent and robot autonomy, between robot plans
and human judgment, between levels of shared control, and between team
roles, handover, and recovery. Table~\ref{tab:mediation-functions}
operationalizes these functions through their information flow,
XR-specific leverage, possible evaluation approach, and relation to the
design dimensions developed in Section~4.}

\changed{The mediation functions and design dimensions serve different purposes.
The mediation functions describe where XR connects human and robot
activity. The design dimensions describe the recurring interface and
evaluation problems that arise when implementing these connections. The
four functions are also interdependent: human
input must be related to robot interpretation, plans must be judged in
context, control relations must remain legible, and team members must be
able to revise robot behavior as roles or circumstances change.}

\subsection{Mediating Between Human Intent and Robot Autonomy}

Robots in collaborative environments are increasingly expected to
respond to high-level human instructions such as ``move this aside,''
``support this part,'' ``prepare the workspace,'' or ``bring the next
tool.'' Such instructions allow users to communicate at the level of
task goals rather than low-level robot motion, but they are frequently
underspecified. The robot must infer referents, select actions, account
for constraints, and determine the appropriate scope of autonomy.

XR can mediate this relation by grounding human input in the shared
environment and externalizing the robot's interpretation before
execution. Speech can be associated with gaze, pointing, or gesture to
identify objects, regions, and directions. Conversely, the interface can
show which referent, goal, target pose, trajectory, or constraint the
robot has inferred. The user can then confirm, refine, or correct that
interpretation in situ. This is particularly important for multimodal
interfaces, where individual signals may remain ambiguous even when
their combination appears plausible
\cite{si2023multimodal,lunding2025arthur}.

The mediation function is therefore bidirectional. XR supplies spatial
context for interpreting human input while also exposing the assumptions
made by robot autonomy. This allows ambiguity to be resolved before an
incorrect interpretation becomes physical robot motion.

\subsection{Mediating Between Robot Plans and Human Judgment}

Robot plans are typically generated with respect to technical objectives
such as reachability, collision avoidance, efficiency, or control
feasibility. Human collaborators evaluate the same plans through
additional criteria, including access, comfort, workload, social
appropriateness, task timing, and the expected actions of other team
members. These criteria are difficult to encode completely because their
importance depends on the current situation.

XR can mediate between robot planning and human judgment by placing
candidate actions and their anticipated consequences in the context in
which they must be evaluated. Relevant representations may include
trajectories and target states, but also contact regions, blocked
workspaces, timing conflicts, proximity to sensitive areas, interference
with another action, or changes in human posture.

In bedside nursing, for example, a collision-free movement of a support
arm must still be assessed relative to the patient's body, the nurse's
access to the treatment site, and the patient's comfort. In collaborative
assembly, a technically feasible handover or arm motion may conflict with
the worker's current subtask, reachable workspace, or next intended
action. XR supports situated judgment by making these relations spatially
inspectable rather than presenting the plan independently of its
consequences.

\subsection{Mediating Between Levels of Shared Control}

Situated human control involves transitions between different forms of
human involvement. A user may initially provide a high-level goal, then
add a constraint, select between alternatives, approve autonomous
execution, directly manipulate a planned path, or interrupt the robot.
Shared-control research has long recognized that autonomy and human
input can be blended or shifted across a task
\cite{dragan2013policy,sheridan1992telerobotics}. The XR-specific
challenge is to make the current control relation and available
transitions legible in the workspace.

An XR mediation layer should therefore communicate not only what the
robot will do, but also how the current action is being controlled. The
interface may need to distinguish autonomous execution, waiting for
approval, following a user-defined constraint, direct trajectory
modification, or safety-limited operation under uncertainty. It should
also show how a new human input will affect the plan.

Making these relations visible can reduce mode confusion and support
deliberate transitions between supervisory and direct control. XR can
additionally expose plans and constraints as spatially manipulable
objects, allowing the user to move from verbal task specification to
deictic correction or direct manipulation without losing the relation to
the physical workspace.

\subsection{Mediating Team Roles, Handover, and Recovery}

Many HRI settings involve several humans with different expertise,
responsibilities, and control rights. A care robot may operate around
nurses, physicians, trainees, patients, and relatives; a teleoperator may
coordinate with another supervisor; and a collaborative robot may work
among several workers. In these settings, not every utterance, gesture,
or gaze cue should be interpreted as a command. The system must account
for who is currently allowed to command, constrain, approve, or interrupt
robot action, and for which robot or subtask that control applies.

XR can mediate team roles by making the active operator, control scope,
pending approvals, and available override mechanisms visible. It can
also support explicit handover by associating control with particular
users, robots, subtasks, or spatial regions. For example, an experienced
worker may temporarily transfer control of one task step to a trainee
while retaining approval or interruption rights. A supervisor may
delegate one manipulator to autonomous execution while directly
controlling another.

Mediation must continue during execution because an accepted plan may
become inappropriate when people move, uncertainty increases, the
environment changes, or responsibility shifts. XR can support recovery
by communicating plan validity, emerging conflicts, and available
revision options. A user may pause execution, update a referent, add or
remove a constraint, redirect a trajectory, transfer control, or request
an alternative plan. Interruption and recovery thus become normal
elements of teamwork rather than mechanisms reserved only for emergency
stops.

Together, these four mediation functions describe how XR can maintain
alignment between human input, robot interpretation, planned action,
control relations, and changing team situations. They provide the
conceptual basis for the design dimensions and research questions
considered next.

\begin{table*}[t]
\centering
\caption{Operationalization of the four XR mediation functions. The
table specifies information flow, XR-specific leverage, an example
evaluation approach, and related design dimensions. R: robot, H: human, Team: multiple human collaborators.}
\label{tab:mediation-functions}
\scriptsize
\renewcommand{\arraystretch}{1.15}
\begin{tabular}{
    p{0.18\textwidth}
    p{0.07\textwidth}
    p{0.27\textwidth}
    p{0.27\textwidth}
    p{0.15\textwidth}}
\toprule
\textbf{Mediation function} &
\textbf{Flow} &
\textbf{XR-specific leverage} &
\textbf{Operational test} &
\textbf{Related dimensions} \\
\midrule

\textbf{Human intent and robot autonomy} &
H$\rightarrow$R$\rightarrow$H &
Co-register multimodal input with physical referents and externalize
the robot's inferred goal and constraints in situ. &
Introduce ambiguous deictic commands; measure grounding accuracy,
correction success, and correction latency. &
S4.3; S4.4 \\

\textbf{Robot plans and human judgment} &
R$\rightarrow$H$\rightarrow$R &
Anchor planned actions and consequences to people, objects, access
regions, and task context. &
Present feasible but contextually inappropriate plans; measure
detection rate, decision accuracy, and response time. &
S4.1; S4.2; S4.3 \\

\textbf{Levels of shared control} &
H$\leftrightarrow$R &
Make the current control mode and manipulable plan elements visible
in the workspace. Multimodal input for coarse-to-fine robot control. &
Require transitions between instruction, constraint, approval,
correction, and interruption; measure mode comprehension and
intervention performance. &
S4.4; S4.6 \\

\textbf{Team roles, handover, and recovery} &
Team$\leftrightarrow$R &
Associate control scope with users, robots, subtasks, and locations;
make handover and recovery options jointly visible. &
Introduce operator changes or plan invalidation; measure command
attribution, handover time, and successful recovery. &
S4.5; S4.6 \\

\bottomrule
\end{tabular}
\end{table*}

\section{Design Dimensions and Research Agenda}

The mediation perspective leads to a set of design dimensions for XR interfaces that support situated human control. These dimensions describe what must become visible and actionable when human collaborators are expected to judge, shape, authorize, and revise robot autonomy in dynamic environments. They also point to research questions such as: how to represent the current team state, how to support intervention without overloading users, and how to evaluate whether XR improves human influence over robot action?

\changed{The  dimensions were derived by comparing the recurring
coordination problems across the three scenarios and relating them to
the mediation functions discussed above. Joint action possibilities
address the relation between human and robot capabilities;
socio-physical constraints address the relation between technically
feasible and contextually appropriate action; uncertainty and plan
validity address changing or incomplete state information; multimodal
control addresses the translation of human input into robot behavior;
roles and handover address multi-person control; and anticipatory
recovery addresses emerging misalignment during execution.}

\subsection{Joint Action Possibilities}

XR interfaces should represent what the human-robot team can achieve together under current constraints. Existing visualizations often focus on the robot's individual capabilities, such as its reachable workspace, planned trajectory, or navigation goal \cite{gruenefeld2020mindthearm,hauck2025reachvox}. For situated human control, the more relevant question is relational: which actions become possible through the combination of human abilities, robot capabilities, shared objects, timing, and workspace layout?

Joint action possibilities may include reachable handover regions, feasible lifting or stabilization configurations, safe cooperation zones, human access spaces, or coordinated motions between multiple robot arms. In bedside nursing, this could mean visualizing whether a robot can support a patient's leg while preserving the nurse's access to the treatment area. In multi-arm teleoperation, it could mean showing which objects can be handled in parallel, which arm should approach from which side, and where simultaneous motion would create interference.

A central research question is how XR can represent joint action possibilities without turning the workspace into a dense planning display. Future work should investigate compact representations that communicate what the team can do now, what alternatives exist, and which actions require human decision or approval.


\subsection{Socio-Physical Constraints}

Robot action in shared environments is shaped by constraints that combine physical, social, ergonomic, and organizational factors. Collision avoidance is necessary, but many relevant risks emerge before collision occurs. A robot may block access, approach a vulnerable body region, move through a space that feels uncomfortable, demand excessive attention, or violate expectations about appropriate conduct. Research on socially aware navigation highlights the importance of interpersonal distance, comfort, legibility, and contextual appropriateness in robot motion \cite{mavrogiannis2023core,singamaneni2024survey}.

XR can make such socio-physical constraints visible as first-class elements of interaction. Instead of showing a trajectory alone, the interface can relate the trajectory to personal spaces, protected regions, task access zones, preferred approach directions, or areas where human attention is already occupied. In care contexts, these constraints may involve patient dignity, discomfort, treatment access, and caregiver workload. In industrial contexts, they may involve ergonomic posture, shared tool access, and temporal coordination with human subtasks.

The research challenge is to encode such constraints in ways that support rapid judgment without giving soft, context-dependent assessments an unwarranted sense of precision. We need methods for visualizing qualitative appropriateness, borderline actions, and trade-offs between efficiency, safety, comfort, and task continuity.


\subsection{Uncertainty, Plan Validity, and Situational Awareness}

Situated human control depends on the user's ability to understand whether a robot plan remains valid. Plans can become outdated when people move, objects are occluded, sensor readings conflict, the task goal changes, or the robot's interpretation of a command is uncertain. XR interfaces should therefore communicate both planned action and the reliability of the assumptions behind it.

Uncertainty cues should be actionable. A user benefits less from abstract confidence values than from knowing where attention, confirmation, or correction is needed. For example, an XR interface might highlight an uncertain object reference, visually degrade path segments that depend on incomplete sensing, mark regions where tracking quality is low, or indicate that a planned motion requires renewed approval after the environment changes. Prior work on uncertainty visualization emphasizes the tension between informativeness and cognitive load \cite{kamal2021uncertainty}; in XR for HRI, this tension is amplified because uncertainty cues appear in the same space where users coordinate physical action.

This dimension also connects directly to situational awareness. Users must perceive relevant changes, understand their implications for the human-robot task, and anticipate what may happen next. Future work should investigate how XR can support these levels of awareness while keeping robot plans actionable. Important questions include when the system should interrupt the user, when it should quietly update the plan, and when it should request renewed confirmation.


\subsection{Multimodal Control and Correction}

Situated human control requires interaction techniques that match the constraints of collaborative work. Speech is efficient for high-level goals and constraints, such as ``move this aside'' or ``approach from the left.'' Gaze and gesture help establish spatial reference, disambiguate objects, and indicate regions. Direct manipulation allows users to reshape virtual trajectories, adjust target poses, or move constraints in the workspace. Haptics or physical interaction may support interruption, guidance, or close-proximity correction.

The design goal is to match modality to control act. Relevant acts include specifying intent through speech, resolving references through gaze or pointing, defining exclusion zones through gesture, comparing alternative plans through spatial previews, adjusting paths through direct manipulation, approving execution explicitly, interrupting action through voice or gesture, and handing over authority through visible role assignment.

Research is needed on interaction vocabularies for XR-mediated control. Such vocabularies should allow users to express task-relevant constraints without becoming responsible for low-level robot programming. At the same time, multimodal systems must provide immediate feedback about interpretation: which referent was selected, which constraint was added, how the plan changed, and which control mode is currently active \cite{si2023multimodal,lunding2025arthur}.


\subsection{Roles, Handover, and Accountability}

Many HRI scenarios involve multiple humans with different roles, expertise, and responsibilities. Yet current interaction designs often assume a single active user. Situated human control requires mechanisms for determining who can issue commands, who can approve execution, who can interrupt, and how control can be handed over during a task.

XR interfaces should represent authority as a dynamic part of the team state. This includes the currently authorized operator, the scope and duration of control, pending approvals, and override rights. A lightweight authority model might combine explicit handover, spatial context, speaker or user recognition, gaze direction, and role assignment. The goal is to avoid both overly rigid authentication procedures and unsafe ambiguity about whose input the robot should follow.

This dimension also supports accountability. Approval should communicate what is being approved: a trajectory, a contact point, a target state, an autonomy level, or a bounded time interval. This matters because authority over robot action can shift dynamically during teamwork. XR can help prevent responsibility from becoming diffuse by keeping the current control relation visible.


\subsection{Anticipatory Recovery}

Finally, situated human control should support correction before breakdowns become failures. Collaborative tasks often fail through gradual misalignment: a wrong object reference, delayed handover, unsafe approach direction, outdated plan, or robot action that conflicts with a human's next step. XR can support anticipatory recovery by making emerging conflicts visible early and by providing low-latency ways to revise robot action.

This dimension extends intent visualization from prediction to intervention. The interface can show future conflict states, likely timing mismatches, regions where human and robot action will interfere, or points where uncertainty may invalidate execution. Users can then pause, redirect, constrain, or reassign the robot while maintaining task continuity. In shared-control settings, such recovery may involve moving along the control continuum: from approving an autonomous plan, to specifying a constraint, to directly adjusting a virtual path, to physically interrupting the robot.

Evaluation should therefore go beyond task completion and subjective usability. Relevant measures include plan comprehension, detection of problematic but technically feasible actions, correction accuracy, intervention latency, recovery time, trust calibration, workload, authority clarity, and task continuity. Scenario-based studies can deliberately introduce ambiguous commands, socially inappropriate paths, stale plans, or hidden uncertainty to test whether users notice and respond appropriately. Longitudinal studies are also needed, since users may initially attend to uncertainty and authority cues but later ignore them or over-trust the system.

Taken together, these design dimensions suggest that situated human control is an important design and evaluation goal. The central question is how to make robot action inspectable, adjustable, and accountable under the conditions of real teamwork.

\section{Conclusion}

XR offers a distinctive interface layer for HRI because robot action is
spatial, embodied, and consequential in the physical world. In this
position paper, we argued that XR should be understood as a mediation
layer for situated human control in human-robot teaming. This perspective
extends the design focus beyond displaying robot intent toward supporting
the situated processes through which human collaborators understand,
shape, authorize, interrupt, and recover robot actions.

We grounded this argument in three scenarios: robot-assisted bedside
nursing, multi-arm supervisory control, and collaborative assembly under
divided attention. Across these scenarios, robot autonomy must remain
understandable and adjustable as the task situation changes. People move,
goals shift, sensing remains incomplete, control roles and
responsibilities may change, and plans that were appropriate during
preview may become unsuitable during execution.

We distinguished four mediation functions through which XR can maintain
alignment between human input, robot interpretation, planned action,
levels of shared control, and changing team situations. These functions
describe how XR can ground multimodal input, externalize robot
interpretations, contextualize planned actions, make control relations
legible, and support handover and recovery.

Building on these functions, we identified six design dimensions for
situated human control: joint action possibilities, socio-physical
constraints, uncertainty and plan validity, multimodal control and
correction, roles, handover, and accountability, and anticipatory
recovery. Future systems should be evaluated by whether they can
meaningfully influence what a joint human-robot team does next. Treating XR
as a mediation layer can help make robot autonomy more inspectable,
adjustable, and accountable in dynamic shared environments.

\section{Acknowledgements}
Generative AI (ChatGPT-5.5, OpenAI) was used to assist with language editing, to improve the flow and readability of the manuscript and for image generation. The authors reviewed and assume full responsibility for the content of this article. This research is funded through the project MAPPLE: Multimodal assistive robot platform for nursing tasks to support loads and improve ergonomics. Federal Ministry of Research, Technology and Space, Germany. 2025-2028.

\balance

\bibliographystyle{abbrv-doi}

\bibliography{template}
\end{document}